\documentclass[review]{elsarticle}
\usepackage{lineno,hyperref}
\modulolinenumbers[5]
\usepackage{amsmath}
\usepackage{subfig}
\usepackage{graphicx}
\graphicspath{D:\backslash Masters2022\backslash HalaPaper\backslash LatexTempNew\backslash paperall}
\journal{arXiv}
\usepackage{numcompress}

\begin{document}

\begin{frontmatter}
\title{A systematic perturbative expansion of the solution of the time-independent Gross-Pitaevskii equation}

\author[address]{Ashraf A Abulseoud}
\author[address]{Hala H Alsayad\corref{correspondingauthor}}
\cortext[correspondingauthor]{Corresponding author}
\ead{helsayad@sci.cu.edu.eg}
\author[address]{Tharwat M El-Sherbini}

\address[address]{Department of Physics, Faculty of Science, Cairo University, Giza, Egypt, 12613}

\begin{abstract}
In this article a perturbative solution of the Gross-Pitaevskii(GP) equation in the $D$-dimensional space $R^D$ with a general external potential is studied. The solution describes the condensate wave-function of a gas containing $N$ Bose particles. A criteria for the validity of the perturbative solution is developed. Furthermore expressions for the particle density, the chemical potential, the internal energy and the mean-square radius of the condensate are derived corrected to first order in the coupling constant. The scheme is then applied to obtain the solution of the GP equation in $D=1,2,3$ for external harmonic potentials. It is shown, in each case, that if $N$ exceeds a certain value the solution breaks down.
\end{abstract}


\end{frontmatter}

\section{Introduction}
The long awaited experimental realization of Bose-Einstein condensation(BEC) in 1995 \cite{1995bradley,1995wieman,1995ketterle} has sparked a huge interest in studying the properties of the newly formed phase and its dynamics. The order parameter of this phase transition and the dynamics of BEC are most accurately described near zero temperature for weak inter-particle interactions by a cubic nonlinear Schr\"odinger equation with an external potential, namely the Gross-Pitaevskii(GP) equation \cite{one,pitaevskii1961vortex,three,four,five,six,Yukalov1,pethick,seven}. 
The external potential represents the trapping potential and the nonlinear term describes, within the mean field approximation, the inter-particle interactions which maybe repulsive or attractive. The GP equation has applications in nonlinear phenomena as in nonlinear optics \cite{nonlinear1,Efremidisoptics2009,Exact(2015)wadati,nonlineardrahsraf}, gravitational physics \cite{Moffat,gr2,15124934,rsta2019}, the Josephson effect \cite{josephson1,josephson2} and in condensed matter physics \cite{spintronics}.\\

Due to its diverse applications in various branches of physics, immense theoretical efforts have been dedicated towards solving the GP equation which seems to be non-integrable even in one dimension. This is due to the presence of the external potential. To overcome this difficulty several approaches were developed which may be divided into numerical and analytical methods. The analytical methods can further be divided into exact solutions for particular forms of the external potential and approximate solutions in analytical forms.\\

The first approach is to develop numerical techniques to solve both of the time-dependent and time-independent GP equation. We shall briefly mention some widely used numerical methods, for a detailed review see \cite{NumericaI(2004)IMPminguzzi} and references therein. Several methods were adapted to solve numerically the time-independent GP equation. The variational scheme was used by Bao and Tang \cite{9innumerical} to solve the GP equation for different forms of the external potential. The boundary-eigenvalue method was applied by Edwards and Burnett \cite{edwards} to obtain solutions of the three-dimensional GP equation with spherically symmetric potentials. Adhikari \cite{adhikari1,adhikari2} used the same technique to obtain the ground state solution of the GP equation in two dimensions with a radially symmetric potential. Another efficient technique, based on simulation of the evolution equation in imaginary time through a Wick rotation, was applied by Chiofalo et al \cite{Chiofal} to compute the ground state solution of BEC in a one-dimensional optical lattice with a superimposed harmonic trap.\\

For the time-dependent GP equation several schemes were also developed. Cerimele \cite{18} et al used a synchronized scheme to solve the two-dimensional GP equation with cylindrical symmetry. Time-dependent extensions of the boundary-eigenvalue method were also developed by Adhikari \cite{19} to solve the axially-symmetric two-dimensional GP equation. Muruganandam and Adhikari \cite{20} used a pseudo-spectral method combined with a Runge-Kutta marching scheme to solve the GP equation in three dimensions. Another widely used technique is the time-splitting spectral method used by Bao et al \cite{Baoetal} to solve the GP equation in one, two and three dimensions for weak as well as strong external potentials. A review of other popular methods is also listed in \cite{Bao2013}.\\

The second approach is to develop methods for constructing particular solutions for the GP equation in an analytical form. A family of stationary periodic solutions to the one-dimensional GP equation with a specifically devised periodic potential was obtained for repulsive interactions by Carr et al \cite{23} and Bronski et al \cite{24} and also for attractive potentials \cite{25,26}. Hua-Mei \cite{hua2005} used the mapping method to find exact solutions of the one-dimensional time-dependent GP equation with a magnetic trap. He obtained bright and dark soliton solutions and soliton-like solutions. Similarity transformations were used by Belmonte-Beitia et al \cite{Belmonte1} to map a one-dimensional time-dependent GP equation with specific time-dependent potentials to a one-dimensional stationary nonlinear Schr\"odinger equation (NLS). Various solutions were constructed including breathers, resonant solitons and quasi-periodic solitons. They also applied the same technique to construct periodic solutions for the time-independent GP equation with periodic potentials and space-dependent coupling constant \cite{Belmonte2}. The same approach was applied by Yu \cite{Fajun2013} to obtain families of exact solutions for the time-dependent GP equation in three dimensions with space- and time-dependent coupling constant. Malomed and Stepanyants \cite{Malomed} used the known solutions of the Gardner equation to generate a family of stationary solutions to the GP equation. They also applied the inverse problem to construct potential functions which support solutions relevant to a particular physical system. The modified Kudryashev method was also used by Neirameh \cite{Neirameh} to obtain solutions of the time-dependent GP equation. Liu et al \cite{Yuan} constructed classes of exactly solvable stationary GP equations with variable coefficients, using transformations to a GP equation with a known exact solution.\\

A subclass of the analytical methods is to develop approximate solutions for the GP equation in analytical forms. Belonging to this class is the Thomas-Fermi approximation introduced by Baym and Pethick \cite{Baym}. Another approach is to apply the variational method with trial functions which are usually Gaussian or solitonic to reduce the problem to a system of ordinary differential evolution equations for some free parameters \cite{variationalfetter,variational}.\\

Another approach belonging to this subclass is to apply perturbation theory in the limit of a weak nonlinear interaction. The solution strategy starts with the general solution of an effective one-dimensional GP equation with a harmonic oscillator potential which is then expressed as a series expansion in terms of the solutions of its linear counterpart. The coefficients of the expansion could then be sought for using various methods. Kivshar et al \cite{kivshar} obtained a system of algebraic equations for the coefficients which they solved, within the weak nonlinearity limit, for the ground state mode as well as for the higher order modes. Trallero-Giner et al \cite{Trallero} obtained for the coefficients a system of nonlinear algebraic equations by transforming the GP equation into an integral equation using the Green’s function of the linear operator, whose spectral representation is given in terms of the one-dimensional harmonic oscillator wave-functions. The algebraic system was solved using the iterative method and also using the perturbative method in the weak nonlinearity limit. Shi et al \cite{Yu} approached the same problem using the homotopy analysis method where a homotopy is constructed with an embedding parameter that goes from zero, corresponding to the linear case, to one, corresponding to the nonlinear GP equation. Assuming that in between those extremes the solution varies smoothly as a function of the embedding parameter, a Maclaurin series is constructed with respect to the embedding parameter for the solution. A set of recursive linearized equations is obtained, which is solved using the Galerkin method. Jia-Ren et al \cite{Ren} studied the perturbative solutions of the GP equation in three dimensions with a spherically symmetric harmonic trap. They expressed the results in terms of the Hermite polynomials and compared their results with the numerical solutions of the GP equation.\\

In this paper we will generalize the approach used in reference \cite{Trallero} for an arbitrary external potential and in an arbitrary dimension. The method is then applied to obtain solutions of the GP equation in one dimension, two dimensions with a rotationally-symmetric and three dimensions with a spherically-symmetric potential. The organization of the paper is as follows. In Section 2 the GP equation in the \textit{D}-dimensional space $R^D$ with an arbitrary external potential is transformed into a system of coupled nonlinear integral equations. In Sections 3 and 4 the integral equations are solved successively and the expansion coefficients are determined as well as the corresponding chemical potentials, furthermore the physical interpretation of the normalization condition is discussed. In Section 5 the particle density, the internal energy and the mean-square radius of the condensate are calculated. The method is then applied in Sections 6, 7 and 8 to study the GP equation in one dimension, two dimensions with rotational symmetry and three dimensions with spherical symmetry. In Section 9 we summarize the obtained results and discuss future prospects.

\section{Formulation of the problem}
We consider a gas of $N$ identical Bose particles, each of mass ${m}_{b}$, in an external potential $V_{ext}(\vec{r})$, where $\vec{r}$ is the position vector in the $D$-dimensional space ${R}^{D}$. The gas is assumed to be in a state of thermodynamic equilibrium at zero temperature. The energy of the stationary condensate $ \psi(\vec{r}) $ is given by the energy functional \cite{pethick} page 148: 
\begin{equation}
\label{eqn:2.1}
E[{\psi ^ * },\psi ] = \int {{d^D} r} \,\{ \frac{{{\hbar ^2}}}{{2{m_b}}}|{\vec \nabla _D}\psi (\vec r){|^2} + {V_{ext}}(\vec r)|\psi (\vec r){|^2} + \frac{g_D}{2}{[|\psi (\vec r){|^2}]^2}\}, 
\end{equation}
where ${\vec \nabla _D}$ is the gradient operator in ${R}^{D}$ and ${g_D}$ is the coupling constant which is given for $D=3$ by \cite{pethick} page 114:
\begin{equation}
\label{eqn:2.2}
{g_3} \equiv g = \frac{{4\pi a{\hbar ^2}}}{{{m_b}}}
\end{equation} 
Here \textit{a} is the \textit{s}-wave scattering length. The appropriate expressions for $g_1$ and $g_2$ are given in Section 6 and Section 7 respectively.

If the external potential $V_{ext}(\vec{r})$ is continuous and approaches infinity as $|\vec r| \to \infty $ and if $g_D > 0$ then the energy functional is convex and its minimum gives the ground state of the Bose condensate $\psi (\vec r)$.
In addition $\psi (\vec r)$ is unique up to a global phase which can always be chosen such that it is real valued and positive \cite{dion}. On the other hand if $g_D < 0$ the ground state exists only at low coupling constants and for a limited number of Bosons in $V_{ext}(\vec{r})$ as long as the minimum-energy balances the effective attraction and prevents collapse \cite{NumericaI(2004)IMPminguzzi}.

To obtain the equation of motion for the condensate wave-function $\psi (\vec r)$, we have to minimize the energy functional Eq.(\ref{eqn:2.1}) subject to the constraint that the total number of particles in the system is equal to $N$:
\begin{equation}
\label{eqn:2.3}
\int {{d^D} r} \,|\psi (\vec r){|^2} = N.
\end{equation}
To carry out the minimization we use the method of the Lagrange multipliers. We construct the auxiliary functional:
\begin{equation}
\label{eqn:2.4}
\begin{gathered}
  K[{\psi ^ * },\psi ,\mu \,]\, = \,E[{\psi ^ * },\psi ]\, - \,\mu \,\int {{d^D} r} \,|\psi (\vec r){|^2} \hfill \\
   = \int {{d^D} r} \,\{ \frac{{{\hbar ^2}}}{{2{m_b}}}|{{\vec \nabla }_D}\psi (\vec r){|^2} + [{V_{ext}}(\vec r) - \mu ]|\psi (\vec r){|^2} + \frac{g_D}{2}{[|\psi (\vec r){|^2}]^2}\} , \hfill \\ 
\end{gathered}  
\end{equation} 
where the chemical potential $\mu$ is the Lagrange multiplier which ensures the constancy of the number of particles.
The equation of motion of the condensate wave-function is then
\begin{equation}
\label{eqn:2.5}
\frac{{\delta K[{\psi ^ * },\psi ,\,\mu ]}}{{\delta {\psi ^ * }(\vec r)}} = 0
\end{equation}

Inserting Eq.(\ref{eqn:2.4}) into Eq.(\ref{eqn:2.5}) we obtain the Gross-Pitaevskii(GP) equation:
\begin{equation}
\label{eqn:2.6}
[ - \frac{{{\hbar ^2}}}{{2{m_b}}}{{\vec \nabla }^2}_D + {V_{ext}}(\vec r) - \mu  + g_D|\psi (\vec r){|^2}]\,\psi (\vec r) = 0.
\end{equation}
We shall consider perturbative solutions of the GP equation for the special case $g_D > 0$, and assume that the chemical potential $\mu$ depends on the coupling constant:
\begin{equation}
\label{eqn:2.7}
\mu  = \mu (g_D)
\end{equation}
Furthermore we assume that the problem recommends a characteristic length ${l_c}$ from which we can define a characteristic angular frequency ${\omega _c}$ by: 
\begin{equation}
{\omega _c} = \frac{\hbar }{{{m_b}\,{l_c}^2}}
\end{equation}
Using $l_c$ and ${\omega _c}$ we can define the dimensionless quantities:
\begin{equation}
\label{eqn:2.9}
\vec \xi  = \frac{{\vec r}}{{{l_c}}};\,\,\,\,\,\,\,\tilde \mu  = \frac{\mu }{{\hbar \,{\omega _c}\,}}\,\,\,\,\,\,\,and\,\,\,\,\,\,\,\tilde g_D = l_c^{ - D}\frac{g_D}{{\hbar \,{\omega _c}}}
\end{equation}
and the dimensionless condensate wave-function:
\begin{equation}
\tilde \psi  = l_c^{\frac{D}{2}}\,\psi. 
\end{equation}
In terms of the dimensionless quantities the GP equation takes the form:
\begin{equation}
\label{eqn:2.11}
[ - \frac{1}{2}{{\vec \nabla }^2}_\xi  + {{\tilde V}_{ext}}(\vec \xi ) - \tilde \mu  + \tilde g_D|\tilde \psi (\vec \xi ){|^2}]\,\tilde \psi (\vec \xi ) = 0
\end{equation}
We introduce the linear differential operator:
\begin{equation}
\label{eqn:2.12}
{L_\xi } =  - \frac{1}{2}\,{{\vec \nabla }^2}_\xi \,\, + \,\,{{\tilde V}_{ext}}(\vec \xi ),
\end{equation}
and rearrange Eq.(10) in the form:
\begin{equation}
\label{eqn:2.13}
{L_\xi }\,\tilde \psi (\vec \xi ) = [\,\tilde \mu  - \tilde g|\tilde \psi (\vec \xi ){|^2}]\,\,\tilde \psi (\vec \xi )
\end{equation}
Let \textit{G} be the Green's function of the differential operator ${L_\xi }$:
\begin{equation}
\label{eqn:2.14}
{L_\xi }\,G\,(\,\vec \xi ,\,\vec \eta \,)\, = \,{\delta ^D}(\,\vec \xi  - \vec \eta \,),
\end{equation}
where ${\delta ^D}(\,\vec \xi  - \vec \eta \,)$ is the \textit{D}-dimensional Dirac-delta function. The condensate wave-function $\tilde \psi (\vec \xi )$ satisfies the nonlinear integral equation:
\begin{equation}
\label{eqn:2.15}
\tilde \psi (\vec \xi ) = \int {{d^D}\eta \,\,G\,(\,\vec \xi ,\,\vec \eta \,)\,[\,\tilde \mu  - \tilde g_D|\tilde \psi (\vec \eta ){|^2}]\,\,\tilde \psi (\vec \eta )}.
\end{equation}
We solve the integral equation (\ref{eqn:2.15}) using the small parameter  method \cite{Myskis}. We assume that the condensate wave-function is non-degenerate and expand $\tilde \mu \,(\tilde g)$ and $\tilde \psi (\vec \xi ,\tilde g)$ in a power series in ${\tilde g}$:
\begin{equation}
\label{eqn:2.16}
\tilde \mu \,(\tilde g_D)\, = \sum\limits_{n\, = \,0} {\,{{\tilde g_D}^n}} \,{{\tilde \mu }_n},\,\,\,\,\,\,and\,\,\,\,\,\,\tilde \psi (\vec \xi ,\tilde g_D)\, = \sum\limits_{n\, = \,0} {\,{{\tilde g_D}^n}} \,{{\tilde \psi }^{(n)}}(\vec \xi )
\end{equation}
We substitute Eq.(\ref{eqn:2.16}) into Eq.(\ref{eqn:2.15}) and use the fact that the condensate wave-function is real, we arrive at:
\begin{equation}
\label{eqn:2.17}
\begin{gathered}
  {{\tilde \psi }^{(0)}}(\vec \xi )\, + \,\tilde g_D\,{{\tilde \psi }^{(1)}}(\vec \xi ) + {{\tilde g_D}^2}\,{{\tilde \psi }^{(2)}}(\vec \xi ) + ... = \,{{\tilde \mu }_0}\int {{d^D}\eta \,\,G\,(\,\vec \xi ,\,\vec \eta \,)\,\,{{\tilde \psi }^{(0)}}(\vec \eta )}  \hfill \\
   + \,\tilde g_D\,\int {{d^D}\eta \,\,G\,(\,\vec \xi ,\,\vec \eta \,)\,[{{\tilde \mu }_1}\,\,{{\tilde \psi }^{(0)}}(\vec \eta )\, + } \,{{\tilde \mu }_0}\,\,{{\tilde \psi }^{(1)}}(\vec \eta )\, - \,{{\tilde \psi }^{(0)\,3}}(\vec \eta )\,] \hfill \\
   + {{\tilde g_D}^2}\int {{d^D}\eta \,\,G\,(\,\vec \xi ,\,\vec \eta \,)\,[{{\tilde \mu }_2}\,\,{{\tilde \psi }^{(0)}}(\vec \eta )\, + } \,{{\tilde \mu }_1}\,\,{{\tilde \psi }^{(1)}}(\vec \eta ) + {{\tilde \mu }_0}\,\,{{\tilde \psi }^{(2)}}(\vec \eta )\,\, - \,3\,{{\tilde \psi }^{(0)\,2}}(\vec \eta )\,{{\tilde \psi }^{(1)}}(\vec \eta )\,] \hfill \\
   + ... \hfill \\ 
\end{gathered} 
\end{equation}
Equating the coefficients of similar powers of $\tilde g_D$ on both sides of Eq.(\ref{eqn:2.17}) we obtain the equations:
\begin{equation}
\label{eqn:2.18}
{{\tilde \psi }^{(0)}}(\vec \xi )\, = \,{{\tilde \mu }_0}\int {{d^D}\eta \,\,G\,(\,\vec \xi ,\,\vec \eta \,)\,\,{{\tilde \psi }^{(0)}}(\vec \eta )} 
\end{equation}
\begin{equation}
\label{eqn:2.19}
{{\tilde \psi }^{(1)}}(\vec \xi ) = \,\int {{d^D}\eta \,\,G\,(\,\vec \xi ,\,\vec \eta \,)\,[{{\tilde \mu }_1}\,\,{{\tilde \psi }^{(0)}}(\vec \eta )\, + } \,{{\tilde \mu }_0}\,\,{{\tilde \psi }^{(1)}}(\vec \eta )\, - \,{{\tilde \psi }^{(0)\,3}}(\vec \eta )\,]
\end{equation}
\begin{equation}
\label{eqn:2.20}
\,{{\tilde \psi }^{(2)}}(\vec \xi ) = \,\int {{d^D}\eta \,\,G\,(\,\vec \xi ,\,\vec \eta \,)\,[{{\tilde \mu }_2}\,\,{{\tilde \psi }^{(0)}}(\vec \eta )\, + } \,{{\tilde \mu }_1}\,\,{{\tilde \psi }^{(1)}}(\vec \eta ) + {{\tilde \mu }_0}\,\,{{\tilde \psi }^{(2)}}(\vec \eta )\,\, - \,3\,{{\tilde \psi }^{(0)\,2}}(\vec \eta )\,{{\tilde \psi }^{(1)}}(\vec \eta )\,]
\end{equation}
$$etc...$$\\
We have replaced the GP equation by a system of coupled integral equations in ${{\tilde \psi }^{(0)}}(\vec \xi ),\,\,{{\tilde \psi }^{(1)}}(\vec \xi ),\,\,{{\tilde \psi }^{(2)}}(\vec \xi ),\,...$ and ${{\tilde \mu }_0},\,\,{{\tilde \mu }_1},\,\,{{\tilde \mu }_2},\,...$. The first integral equation is linear in ${{\tilde \psi }^{(0)}}$ and shows that ${{\tilde \psi }^{(0)}}$ is an eigenfunction of the linear operator ${L_\xi }$ corresponding to the eigenvalue ${{\tilde \mu }_0}$. Solving this equation we obtain ${{\tilde \psi }^{(0)}}$ and ${{\tilde \mu }_0}$. Substituting these quantities back into Eq.(\ref{eqn:2.18}) we obtain an inhomogeneous linear integral equation in ${{\tilde \psi }^{(1)}}$ and ${{\tilde \mu }_1}$ which can easily be solved. Continuing this process we can, in principle, obtain ${{\tilde \psi }^{(2)}}(\vec \xi ),\,\,{{\tilde \psi }^{(3)}}(\vec \xi ),...$ and ${{\tilde \mu }_2},\,\,{{\tilde \mu }_3},...$. This gives a perturbative solution of the GP equation to any desired order of ${\tilde g_D}$.

\section{Solution of the system of integral equations}
To obtain the Green's function associated with the differential operator ${L_\xi }$ we have to solve the eigenvalue problem:
\begin{equation}
\label{eqn:3.1}
{L_\xi }\,\chi \,(\,\vec \xi \,)\, = \,\lambda \chi \,(\,\vec \xi \,).
\end{equation}
For a large class of potentials, ${V_{ext}}$, \cite{Titchmarsh1, Titchmarsh2}, the eigenvalue problem \ref{eqn:3.1} has a complete set of orthonormal eigenfunctions:
\begin{equation}
\label{eqn:3.2}
\{ {\chi _{\alpha (n)}}\,(\,\vec \xi \,)\}, 
\end{equation} 
where $\alpha (n)$ is a complete set of indices characterizing the eigenvalues and the eigenfunctions.
\begin{equation}
\label{eqn:3.3}
\int {{d^D}\xi \,\,} {\chi ^*}_{\alpha (n)}\,(\,\vec \xi \,)\,{\chi _{\alpha (k)}}\,(\,\vec \xi \,)\, = \,{\delta _{\alpha (n),\,\alpha (k)}}
\end{equation}
\begin{equation}
\label{eqn:3.4}
\sum\limits_n {{\chi ^*}_{\alpha (n)}\,(\,\vec \eta \,)\,{\chi _{\alpha (k)}}\,(\,\vec \xi \,)\,\, = \,{\delta ^D}\,(} \vec \xi \, - \vec \eta )
\end{equation}
We will label the eigenvalues according to their magnitudes
\begin{equation}
0 < {\lambda _{\,\alpha \,(0)}}\, < {\lambda _{\,\alpha \,(1)}}\, \leq {\lambda _{\,\alpha \,(2)}} \leq ... \leq \,{\lambda _{\,\alpha \,(n)}} \leq ...
\end{equation}
The Green's function associated with the differential operator can be expressed as:
\begin{equation}
G\,(\,\vec \xi ,\,\vec \eta \,)\, = \,\sum\limits_n {\frac{1}{{{\lambda _{\,\alpha \,(n)}}}}{\chi ^*}_{\alpha (n)}\,(\,\vec \eta \,)\,{\chi _{\alpha (n)}}\,(\,\vec \xi \,).\,} 
\end{equation}
The integral equation (\ref{eqn:2.15}) has the form:
$$\tilde \psi (\vec \xi ) = \int {{d^D}\eta \,\,G\,(\,\vec \xi ,\,\vec \eta \,)\,\varphi \,(\vec \eta \,),}$$
where 
$$\varphi \,(\vec \xi \,) = [\,\tilde \mu  - \tilde g|\tilde \psi (\vec \xi ){|^2}]\,\,\tilde \psi (\vec \xi ).$$
Since the condensate wave-function $\tilde \psi (\vec \xi )$ is assumed to be square integrable and vanishes as $|\vec \xi |\, \to \,0$ the function $\varphi \,(\vec \xi \,)$ is also square integrable and vanishes as $|\vec \xi |\, \to \,0$. Furthermore the kernel, $G\,(\,\vec \xi ,\,\vec \eta \,)$, of the integral equation is Hermitian and square integrable. Therefore, according to the Hilbert-Schmidt theorem \cite{integral}, the function $\tilde \psi (\vec \xi )$ can be expanded in terms of the eigenfunctions (\ref{eqn:3.2}).
In particular, we can expand the functions ${{\tilde \psi }^{(n)}}(\vec \xi )\,$ in terms of this set:
\begin{equation}
\label{eqn:3.7}
{{\tilde \psi }^{(0)}}(\vec \xi )\, = \,\sum\limits_{k = 0} {C_{\alpha \,(k)}^{(0)}} \,{\chi _{\alpha (k)}}\,(\,\vec \xi \,)
\end{equation}
\begin{equation}
\label{eqn:3.8}
{{\tilde \psi }^{(n)}}(\vec \xi )\, = \,C_{\alpha \,(0)}^{(0)}\,\sum\limits_{k = 0} {C_{\alpha \,(k)}^{(n)}} \,{\chi _{\alpha (k)}}\,(\,\vec \xi \,),\,\,\,\,\,n = 1,2,...
\end{equation}
Substituting Eq.(\ref{eqn:3.7}) into Eq.(\ref{eqn:2.18}) and using the orthogonality relation (\ref{eqn:3.3}) we obtain
$$(\,{\lambda _{\,\alpha \,(k)}}\, - \,{{\tilde \mu }_0})\,C_{\alpha \,(k)}^{(0)}\, = \,0$$
This equation shows that $\tilde \mu _0$ is equal to one of the eigenvalues of the linear operator $L_\xi $, say $k = m$. Then 
\begin{equation}
\label{eqn:3.9}
\begin{gathered}
  {{\tilde \mu }_0}\, = \,{\lambda _{\,\alpha \,(m)}},\,\,and\,\,\,C_{\alpha \,(m)}^{(0)}\, \ne \,0 \hfill \\
  C_{\alpha \,(k)}^{(0)}\, = \,0,\,\,for\,\,k \ne m \hfill \\ 
\end{gathered} 
\end{equation}
This solution is of particular importance since it represents the case in which most of the particles condense in the state ${\chi _{\alpha (m)}}$ of the trapping potential. At $T = 0$ all Bose particles condense in the ground state so that $m = 0$. Thus 
\begin{equation}
\label{eqn:3.10}
C_{\alpha \,(k)}^{(0)} = \,C_{\alpha \,(0)}^{(0)}\,{\delta _{k,0}}.
\end{equation}
Inserting this back into Eqs.(\ref{eqn:3.7}) and (\ref{eqn:3.9}), we obtain
\begin{equation}
\label{eqn:3.11}
{{\tilde \psi }^{(0)}}(\vec \xi )\, = \,C_{\alpha \,(0)}^{(0)}\,{\chi _{\alpha (0)}}\,(\,\vec \xi \,)
\end{equation}
\begin{equation}
\label{eqn:3.12}
{{\tilde \mu }_0}\, = \,{\lambda _{\,\alpha \,(0)}}
\end{equation}
We next consider the integral equation (\ref{eqn:2.19}). substituting Eq.(\ref{eqn:3.8}) with $n = 1$ into Eq.(\ref{eqn:2.19}) we obtain:
\begin{equation}
\label{eqn:3.13}
\begin{gathered}
  C_{\alpha \,(0)}^{(1)}\,{\chi _{\alpha (0)}}\,(\,\vec \xi \,)\, + \,\sum\limits_{k \ne 0} {C_{\alpha \,(k)}^{(1)}} \,{\chi _{\alpha (k)}}\,(\,\vec \xi \,)\, = \,\frac{1}{{{\lambda _{\,\alpha \,(0)}}}}\,[\,{{\tilde \mu }_1} - \,C_{\alpha \,(0)}^{(0)\,2}\,{b_{\alpha \,(0)}}\, + \,{\lambda _{\,\alpha \,(0)}}\,C_{\alpha \,(0)}^{(1)\,}]\,{\chi _{\alpha (0)}}\,(\,\vec \xi \,)\, \hfill \\
   + \,\sum\limits_{k \ne 0} {\frac{1}{{{\lambda _{\,\alpha \,(k)}}}}\,[\, - \,C_{\alpha \,(0)}^{(0)\,2}\,{b_{\alpha \,(k)}}\, + \,{\lambda _{\,\alpha \,(0)}}\,C_{\alpha \,(k)}^{(1)\,}]\,{\chi _{\alpha (k)}}\,(\,\vec \xi \,)\,,}  \hfill \\ 
\end{gathered} 
\end{equation}
where 
\begin{equation}
\label{eqn:3.14}
{b_{\alpha \,(k)}} = \int {{d^D}\eta \,\,} {\chi ^*}_{\alpha (k)}\,(\,\vec \xi \,)\,\chi _{\alpha (0)}^3\,(\,\vec \xi \,).
\end{equation}
Equating the coefficients of the corresponding eigenfunctions on both sides gives:
\begin{equation}
\label{eqn:3.15}
{{\tilde \mu }_1}\, = \,C_{\alpha \,(0)}^{(0)\,2}\,\,{b_{\alpha \,(0)}},\,
\end{equation}
and 
\begin{equation}
\label{eqn:3.16}
C_{\alpha \,(k)}^{(1)\,}\, = \, - \,C_{\alpha \,(0)}^{(0)\,2}\,{T_{\alpha \,(k)}};\,\,\,\,\,\,\,k \ne 0,
\end{equation}
with
\begin{equation}
\label{eqn:3.17}
{T_{\alpha \,(k)}}\, = \,\,\frac{{{b_{\alpha \,(k)}}}}{{{\lambda _{\,\alpha \,(k)}} - {\lambda _{\,\alpha \,(0)}}}};\,\,\,\,\,\,\,k \ne 0
\end{equation}
To obtain ${{\tilde \psi }^{(2)}}$ we have to solve the integral equation(\ref{eqn:2.20}). In this case we should take into account the change in the coupling constant $g$, due to higher order inter-particles scattering processes. This means that we have to go beyond the \textit{s}-wave expression in Eq.(\ref{eqn:2.2}). Although we will mainly be concerned with the condensate wave-function corrected to first order in \textit{g}, we will go one step further and calculate the second order term to get a deeper look at the expansion and to explain how the normalization condition (\ref{eqn:2.3}) is satisfied to all orders of the coupling constant.
Substituting Eqs(\ref{eqn:3.8}), with $n = 1$ and $n = 2$ into Eq.(\ref{eqn:2.20}) we obtain
\begin{equation}
\label{eqn:3.18}
\begin{gathered}
  C_{\alpha \,(0)}^{(2)}\,{\chi _{\alpha (0)}}\,(\,\vec \xi \,)\, + \,\sum\limits_{k \ne 0} {C_{\alpha \,(k)}^{(2)}} \,{\chi _{\alpha (k)}}\,(\,\vec \xi \,)\, =  \hfill \\
  \,\frac{1}{{{\lambda _{\,\alpha \,(0)}}}}\,[\,{{\tilde \mu }_2} - \,3\,C_{\alpha \,(0)}^{(0)\,2}\,\,\sum\limits_{l = 0} {{D_{\alpha \,(0),\alpha \,(l)}}\,C_{\alpha \,(l)}^{(1)\,}} \, + \,{{\tilde \mu }_1}\,C_{\alpha \,(0)}^{(1)\,} + {{\tilde \mu }_0}\,C_{\alpha \,(0)}^{(2)\,}]\,{\chi _{\alpha (0)}}\,(\,\vec \xi \,)\, \hfill \\
   + \,\sum\limits_{k \ne 0} {\frac{1}{{{\lambda _{\,\alpha \,(k)}}}}\,[\, - 3\,C_{\alpha \,(0)}^{(0)\,2}\,\,\sum\limits_{l = 0} {{D_{\alpha \,(k),\alpha \,(l)}}\,C_{\alpha \,(l)}^{(1)\,}} \, + \,{{\tilde \mu }_1}\,C_{\alpha \,(k)}^{(1)\,} + {{\tilde \mu }_0}\,C_{\alpha \,(k)}^{(2)\,}]\,{\chi _{\alpha (k)}}\,(\,\vec \xi \,)\,,}  \hfill \\ 
\end{gathered}  
\end{equation}
where 
\begin{equation}
\label{eqn:3.19}
{D_{\alpha \,(k),\,\alpha \,(l)}} = \int {{d^D}\eta \,\,} {\chi ^*}_{\alpha (k)}\,(\,\vec \xi \,)\,\chi _{\alpha (0)}^2\,(\,\vec \xi \,){\chi _{\alpha \,(l)}}\,(\,\vec \xi \,).
\end{equation}
Equating the coefficients of the corresponding eigenfunctions on both sides of Eq.(\ref{eqn:3.17}) we obtain:
\begin{equation}
\label{eqn:3.20}
{{\tilde \mu }_2} = 3\,C_{\alpha \,(0)}^{(0)\,2}\,\,\sum\limits_{l = 0} {{D_{\alpha \,(0),\alpha \,(l)}}\,C_{\alpha \,(l)}^{(1)\,}} \, - \,{{\tilde \mu }_1}\,C_{\alpha \,(0)}^{(1)\,},
\end{equation}
and 
\begin{equation}
\label{eqn:3.21}
C_{\alpha \,(k)}^{(2)}\, = \,\,\frac{1}{{{\lambda _{\,\alpha \,(k)}}}}\,[\, - 3\,C_{\alpha \,(0)}^{(0)\,2}\,\,\sum\limits_{l = 0} {{D_{\alpha \,(k),\alpha \,(l)}}\,C_{\alpha \,(l)}^{(1)\,}} \, + \,{{\tilde \mu }_1}\,C_{\alpha \,(k)}^{(1)\,} + {{\tilde \mu }_0}\,C_{\alpha \,(k)}^{(2)\,}],\,\,\,\,\,\,\,k \ne 0.
\end{equation}
Using the results 
$${D_{\alpha \,(0),\alpha \,(l)}}\, = \,\,b_{\alpha \,(k)}^*;\,\,\,\,\,\,b_{\alpha \,(0)}^* = {b_{\alpha \,(0)}},$$
we can express the chemical potential as:
\begin{equation}
\label{eqn:3.22}
{{\tilde \mu }_2} = 2\,C_{\alpha \,(0)}^{(0)\,2}\,C_{\alpha \,(0)}^{(1)\,}\,{b_{\alpha \,(0)}} - 3\,\,C_{\alpha \,(0)}^{(0)\,4}\,\,\sum\limits_{l \ne 0} {b_{\alpha \,(l)}^*{T_{\alpha \,(l)}}\,} \,
\end{equation}
Substituting the expressions for $C_{\alpha \,(k)}^{(1)\,}$, ${{\tilde \mu }_1}$ and ${{\tilde \mu }_0}\,$ we obtain:
\begin{equation}
\label{eqn:3.23}
\begin{gathered}
  C_{\alpha \,(k)}^{(2)}\, = \,\, - 3\frac{{C_{\alpha \,(0)}^{(0)\,2}}}{{{\lambda _{\,\alpha \,(k)}} - \,\,{\lambda _{\,\alpha \,(0)}}}}\,[\,\,C_{\alpha \,(0)}^{(1)\,}\,{b_{\alpha \,(k)}} - \,\,\frac{1}{3}C_{\alpha \,(0)}^{(0)\,2}\,{b_{\alpha \,(0)}}\,{T_{\alpha \,(k)}} \hfill \\
   - \,C_{\alpha \,(0)}^{(0)\,2}\,\,\sum\limits_{l \ne 0} {{D_{\alpha \,(k),\alpha \,(l)}}\,{T_{\alpha \,(l)}}} \,],\,\,\,\,\,\,\,k \ne 0 \hfill \\ 
\end{gathered} 
\end{equation}
Eqs(\ref{eqn:3.10}),(\ref{eqn:3.16}) and (\ref{eqn:3.23}) give the expansion coefficients in terms of the, in principle, known quantities ${b_{\alpha \,(k)}}$ and ${{D_{\alpha \,(k),\alpha \,(l)}}}$. We still have to determine $C_{\alpha \,(0)}^{(0)\,}$, $C_{\alpha \,(0)}^{(1)\,}$ and $C_{\alpha \,(0)}^{(2)\,}$.
Substituting Eqs(\ref{eqn:3.7}) and (\ref{eqn:3.8}) with $n = 1, 2$ into Eq.(\ref{eqn:2.16}) we obtain:
\begin{equation}
\label{eqn:3.24}
\begin{gathered}
  \tilde \psi (\vec \xi )\, = \,C_{\alpha (0)}^{(0)}\,\{ \,[1 + \,\tilde g_D\,C_{\alpha (0)}^{(1)} + \,{{\tilde g_D}^2}\,C_{\alpha (0)}^{(2)}\,]\,{\chi _{\alpha (0)}}(\vec \xi ) + \, \hfill \\
  \sum\limits_{k \ne 0} {\,[\,\tilde g_D\,C_{\alpha (k)}^{(1)} + \,{{\tilde g_D}^2}\,C_{\alpha (k)}^{(2)}\,]\,{\chi _{\alpha (k)}}(\vec \xi )} \,\}  + \,O({{\tilde g_D}^3}). \hfill \\ 
\end{gathered} 
\end{equation}

\section{Determination of the constants $C_{\alpha (0)}^{(0)}$, $C_{\alpha (0)}^{(1)}$, $C_{\alpha (0)}^{(2)}$}
The constants $C_{\alpha (0)}^{(0)}$, $C_{\alpha (0)}^{(1)}$, $C_{\alpha (0)}^{(2)}$ describe the effect of the inter-particle interactions on the ground state and therefore may depend on the coupling constant ${\tilde g_D}$. On the other hand, $C_{\alpha (k)}^{(1)}$ and $C_{\alpha (k)}^{(2)}$ for $k \ne 0$, are completely determined by Eq.(\ref{eqn:3.14}) and Eq.(\ref{eqn:3.19}), and are independent of ${\tilde g_D}$. 
We determine the unknown constants using the normalization condition:
\begin{equation}
\label{eqn:4.1}
\int {{d^D}\xi } \,\,{{\tilde \psi }^*}(\vec \xi )\,\tilde \psi (\vec \xi ) = N
\end{equation}
Inserting Eq.(\ref{eqn:3.24}) into Eq.(\ref{eqn:4.1}) we obtain:
\begin{equation}
\label{eqn:4.2}
C_{\alpha (0)}^{(0)\,2}\,\{ \,[1 + \,\tilde g_D\,C_{\alpha (0)}^{(1)} + \,{{\tilde g_D}^2}\,C_{\alpha (0)}^{(2)}\,]{\,^2} + \,\sum\limits_{k \ne 0} {\,|\,\tilde g_D\,C_{\alpha (k)}^{(1)} + \,{{\tilde g_D}^2}\,C_{\alpha (k)}^{(2)}\,{|^2}} \,\}  = \,N
\end{equation}
We introduce the quantities:
\begin{equation}
\label{eqn:4.3}
\begin{gathered}
  C_{\alpha (0)}^{(0)\,4}\,{S^{(n)}} = \,\sum\limits_{k \ne 0} {\,|\,C_{\alpha (k)}^{(n)}\,{|^2},\,\,n = 1,2} \, \hfill \\
  C_{\alpha (0)}^{(0)\,4}\,{S^{(1,2)}} = \,\sum\limits_{k \ne 0} {\,[\,C_{\alpha (k)}^{(1)\,*}\,C_{\alpha (k)}^{(2)}\, + \,C_{\alpha (k)}^{(1)}\,C_{\alpha (k)}^{(2)\,*}]} \, \hfill \\ 
\end{gathered}
\end{equation}
 
and rewrite Eq.(\ref{eqn:4.2}) in the form:
\begin{equation}
\label{eqn:4.4}
\begin{gathered}
  C_{\alpha (0)}^{(0)\,2}\,\{ \,1 + \,2\,\tilde g_D\,C_{\alpha (0)}^{(1)} + \,2\,{{\tilde g_D}^2}\,C_{\alpha (0)}^{(2)}\, + 2\,{{\tilde g_D}^3}\,C_{\alpha (0)}^{(1)}\,C_{\alpha (0)}^{(2)}\, + \,{{\tilde g_D}^2}\,C_{\alpha (0)}^{(1)\,2}\, + \,{{\tilde g_D}^4}\,C_{\alpha (0)}^{(2)\,2}\, \hfill \\
   + \,C_{\alpha (0)}^{(0)\,4}\,[\,{{\tilde g_D}^2}\,{S^{(1)}} + \,{{\tilde g_D}^3}\,{S^{(1,2)}}\, + \,{{\tilde g_D}^4}\,{S^{(2)}}]\,\}  = \,N \hfill \\ 
\end{gathered} 
\end{equation}
Taking the limit $\tilde g_D \to 0$, we obtain:
\begin{equation}
\label{eqn:4.5}
C_{\alpha (0)}^{(0)} = \,\sqrt N 
\end{equation}
Inserting this back into Eq.(\ref{eqn:4.4}) leads to
\begin{equation}
\label{eqn:4.6}
\begin{gathered}
  2\,C_{\alpha (0)}^{(1)} + \,2\,\tilde g_D\,C_{\alpha (0)}^{(2)}\, + 2\,{{\tilde g_D}^2}\,C_{\alpha (0)}^{(1)}\,C_{\alpha (0)}^{(2)}\, + \,\tilde g_D\,C_{\alpha (0)}^{(1)\,2}\, + \,{{\tilde g_D}^3}\,C_{\alpha (0)}^{(2)\,2}\, \hfill \\
   + \,{N^2}\,[\,\tilde g_D\,{S^{(1)}} + \,{{\tilde g_D}^2}\,{S^{(1,2)}}\, + \,{{\tilde g_D}^3}\,{S^{(2)}}]\, = \,0 \hfill \\ 
\end{gathered} 
\end{equation}
We choose
\begin{equation}
\label{eqn:4.7}
C_{\alpha (0)}^{(1)} =  - \frac{1}{2}{N^2}\,\tilde g_D\,{S^{(1)}}.
\end{equation}
Substituting this back into Eq.(\ref{eqn:4.6}) we obtain
\begin{equation}
\label{eqn:4.8}
\begin{gathered}
  2\,C_{\alpha (0)}^{(2)}\, - \,{{\tilde g_D}^2}\,{N^2}{S^{(1)}}C_{\alpha (0)}^{(2)}\, + \frac{1}{4}\,{{\tilde g_D}^2}\,{N^4}\,{S^{(1)\,2}}\, + \,{{\tilde g_D}^2}\,C_{\alpha (0)}^{(2)\,2}\, \hfill \\
   + \,{N^2}\,[\,\tilde g_D\,{S^{(1,2)}}\, + \,{{\tilde g_D}^2}\,{S^{(2)}}]\, = \,0 \hfill \\ 
\end{gathered} 
\end{equation}
We choose 
\begin{equation}
\label{eqn:4.9}
C_{\alpha (0)}^{(2)} =  - \frac{1}{2}{N^2}\,\tilde g_D\,{S^{(1,2)}}
\end{equation}
The remaining terms in Eq.(\ref{eqn:4.8})are of higher order and can be safely neglected since they determine the constants $C_{\alpha (0)}^{(3)}$ and $C_{\alpha (0)}^{(4)}$ which we have ignored. Thus 
\begin{equation}
\label{eqn:4.10}
\tilde \psi (\vec \xi )\, = \,\sqrt N \{ \,[1 - \frac{1}{2}\,{(\tilde g_D\,N)^2}{S^{(1)}}]\,{\chi _{\alpha (0)}}(\vec \xi ) + \,\sum\limits_{k \ne 0} {\,[\,\tilde g_D\,C_{\alpha (k)}^{(1)} + \,{{\tilde g_D}^2}\,C_{\alpha (k)}^{(2)}\,]\,{\chi _{\alpha (k)}}(\vec \xi )} \,\}  + \,O({{\tilde g_D}^3}) 
\end{equation}
The normalization condition, corrected to second order in ${\tilde g}$, now reads 
\begin{equation}
\label{eqn:4.11}
\{ \,[1 - {(\tilde g_D\,N)^2}{S^{(1)}}]\, + {(\tilde g_D\,N)^2}{S^{(1)}}\}  + \,O({{\tilde g_D}^3}) = 1
\end{equation}
Eq.(\ref{eqn:4.11}) suggests that we interpret 
$$1 - {(\tilde g_D\,N)^2}{S^{(1)}},$$
as the fraction, $\frac{{{N_0}}}{N}$, of Bose particles in the ground state and
$${(\tilde g_D\,N)^2}{S^{(1)}},$$
as the fraction, $\frac{{{N_\varepsilon }}}{N}$, of particles tunneled to the excited states due to the inter-particle interactions. To accept this interpretation we should have: 
\begin{equation}
\label{eqn:4.12}
{(\tilde g_D\,N)^2}{S^{(1)}} \ll \,1.
\end{equation}
The breakdown of this condition indicates the breakdown of the perturbative solution.

\section{Some physical parameters of the system}
The chemical potential to first order in ${\tilde g_D}$ is given by:
\begin{equation}
\label{eqn:5.1}
\tilde \mu \, = \,{\lambda _{\alpha (0)}} + \tilde g_D\,N\,{b_{\alpha (0)}} + O({{\tilde g_D}^2})
\end{equation}
If you are interested in the second-order correction, ${{\tilde \mu }_2}$, then we obtain from Eq.(\ref{eqn:3.20}) 
\begin{equation}
\label{eqn:5.2}
{{\tilde \mu }_2}\, = \, - 3{N^2}\,\sum\limits_{k \ne 0} {b_{\alpha (k)}^*\,{T_{\alpha (k)}}} 
\end{equation}
Here we have neglected the second term in Eq.(\ref{eqn:3.20}) since it contains a ${\tilde g}$ in the factor $C_{\alpha (0)}^{(1)}$.\\

Another interesting quantity is the particle density 
\begin{equation}
\label{eqn:5.3}
n(\vec \xi )\, = \,|\tilde \psi (\vec \xi ){|^2}
\end{equation}
Using Eq.(\ref{eqn:4.10}) we obtain, after neglecting terms of order ${{\tilde g}^2}$:
\begin{equation}
\label{eqn:5.4}
n(\vec \xi )\, = \,N\{ \,\chi _{\alpha (0)}^2(\vec \xi ) + \,\tilde g_D\,\sum\limits_{k \ne 0} {\,[\,C_{\alpha (k)}^{(1)\,*}\,\chi _{\alpha (k)}^*(\vec \xi )\, + \,C_{\alpha (k)}^{(1)\,}\,{\chi _{\alpha (k)}}(\vec \xi )\,]\,{\chi _{\alpha (0)}}(\vec \xi )} \,\}  + \,O({{\tilde g_D}^2})
\end{equation}
The internal energy of the condensate $U(N,\,{\omega _c})$ is identified with the energy functional Eq.(\ref{eqn:2.1}) evaluated at the actual configuration of the system, given by the condensate wave-function satisfying the GP equation. We first integrate the first term of Eq.(\ref{eqn:2.1}) by parts and apply Gauss' theorem to obtain:
\[\,E[{\psi ^ * },\psi ] = \,\int {{d^D}r} \,{\psi ^*}(\vec r)\,[\, - \frac{{{\hbar ^2}}}{{2{m_b}}}\vec \nabla _D^2\,\psi (\vec r) + {V_{ext}}(\vec r) + \frac{g_D}{2}|\psi (\vec r){|^2}]\,\psi (\vec r).\]
We next use the GP equation to simplify this expression and then express the result in the dimensionless variables. This gives:
\begin{equation}
\label{eqn:5.5}
\tilde U(N,\,{\omega _c}) = \,\int {{d^D}\xi \,} [\,\tilde \mu  - \frac{{\tilde g_D}}{2}|\tilde \psi (\vec \xi ){|^2}]\,\,|\tilde \psi (\vec \xi ){|^2}
\end{equation}
Substituting Eqs.(\ref{eqn:4.10}) and (\ref{eqn:5.5}) we obtain:
\begin{equation}
\label{eqn:5.6}
\tilde U(N,\,{\omega _c}) = \,N\,[{\lambda _{\alpha (0)}} + \frac{1}{2}\tilde g_D\,N\,{b_{\alpha (0)}}] + O({{\tilde g_D}^2}).
\end{equation}
As a check, we calculate the chemical potential using the thermodynamic relation:
\[\tilde \mu  = {(\frac{{\partial \tilde U}}{{\partial N}})_{{\omega _c}}}.\]
This reproduces Eq.(\ref{eqn:5.1}).
\\The shape of the atomic cloud can be characterized by the mean-square radius of the condensate
\begin{equation}
\label{eqn:5.7}
{r_0}\, = \,\sqrt { < {r^2} > } 
\end{equation}
The dimensionless mean-square radius is then given by:
\begin{equation}
\label{eqn:5.8}
{\xi _0} = \,\sqrt {\int {{d^D}\xi \,} {\xi ^2}n(\vec \xi )} 
\end{equation}
Using Eq.(\ref{eqn:5.4}) and keeping only terms linear in ${\tilde g}$ we obtain:
\begin{equation}
\label{eqn:5.9}
\xi _0^2\, = \,N\,\{ {M_{\alpha (0),\alpha (0)}} - \tilde g_D\,N\sum\limits_{k \ne 0} {[T_{\alpha (k)}^*{M_{\alpha (k),\alpha (0)}} + \,{T_{\alpha (k)}}M_{\alpha (k),\alpha (0)}^*]} \}  + O({{\tilde g_D}^2}),
\end{equation}
where
\begin{equation}
\label{eqn:5.10}
{M_{\alpha (k),\alpha (l)}} = \,\int {{d^D}\xi \,{\xi ^2}\,\chi _{\alpha (k)}^*(\vec \xi )} \,{\chi _{\alpha (l)}}(\vec \xi ).\,
\end{equation}
Taking the square root of Eq.(\ref{eqn:5.9}) we arrive at:
\begin{equation}
\label{eqn:5.11}
{\xi _0} = \,\sqrt {N{M_{\alpha (0),\alpha (0)}}} \,\{ 1\, - \,\frac{{\tilde g_D\,N}}{{2{M_{\alpha (0),\alpha (0)}}}}\,\sum\limits_{k \ne 0} {[T_{\alpha (k)}^*{M_{\alpha (k),\alpha (0)}} + \,{T_{\alpha (k)}}M_{\alpha (k),\alpha (0)}^*]} \,\} 
\end{equation}

\section{The one-dimensional harmonic potential}
We now consider particular forms of the external potential and begin with the one-dimensional harmonic potential
\begin{equation}
\label{eqn:6.1}
{V_{ext}}(x) = \frac{1}{2}{m_b}\,{\omega ^2}{x^2}
\end{equation}
The experimental realization of the $D=1$ condensate is carried out by trapping the Bose gas in the potential
\[{V_{ext}} = \frac{1}{2}{m_b}\,(\omega _x^2{x^2} + \omega _y^2{y^2} + \omega _z^2{z^2}).\] 
By increasing the frequencies $\omega _y$ and $\omega _z$ the confinement of the condensate along these axes is increased and the dynamics of the system in these directions are restricted to the zero-point oscillation. The $x$-component of the wave-function can then be factored, so that the system can approximately be described by the one-dimensional GP-equation. In this case the one-dimensional coupling constant is given by \cite{dion}
\[{g_1} = \,g\,\sqrt {\frac{{{m_b}\,{\omega _y}}}{{2\,\pi \,\hbar }}} \,\sqrt {\frac{{{m_b}\,{\omega _z}}}{{2\,\pi \,\hbar }}} .\]
Using Eq.(\ref{eqn:2.2}) we obtain:
\begin{equation}
\label{eqn:g1D}
{{\tilde g}_1} = \,\frac{{2\,a}}{{{l_c}}}\sqrt {{\gamma _y}\,{\gamma _z}} ,
\end{equation}
where,
\[{l_c} = \,\,\sqrt {\frac{\hbar }{{{m_b}\,\omega }}} ;\,\,\,\,\,\,\,{\gamma _y} = \frac{{{\omega _y}}}{{{\omega _x}}};\,\,\,\,\,\,\,{\gamma _z} = \frac{{{\omega _z}}}{{{\omega _x}}}.\]

With the choice of the potential in Eq.(\ref{eqn:6.1}), Eq.(\ref{eqn:3.1}) reduces to the eigenvalue problem for the one-dimensional harmonic oscillator. The eigenfunctions and eigenvalues are given, respectively, by:
\begin{equation}
\label{eqn:6.2}
{\chi _n}(\xi )\, = \,\frac{1}{{\sqrt {{2^n}\,n!\,\sqrt \pi  } }}\,\,\exp ( - \frac{1}{2}{\xi ^2})\,{H_n}(\xi ),\,\,\,\,\,\,\,\,\, - \infty  < \xi  < \infty ,
\end{equation}
and 
\begin{equation}
\label{eqn:6.3}
{\lambda _n}\, = \,n + \frac{1}{2}.
\end{equation}
The Hermite polynomials of degree \textit{n}, ${H_n}$, \cite{Bellspecialfunctions,Lebedevspecialfunctions} has the series representation
\begin{equation}
\label{eqn:6.4}
{H_n}(x)\, = \,\sum\limits_{r\, = \,0}^{[n/2]} {{{( - 1)}^r}} \,\frac{{n!}}{{r!\,(n - 2r)!}}\,{(2x)^{n - 2r}},
\end{equation}
where $[n/2]$ is the integral part of $n/2$.
The ground state wave-function
\begin{equation}
\label{eqn:6.5}
{\chi _0}(\xi )\, = {(\frac{1}{\pi })^{1/4}}\,\exp ( - \frac{1}{2}{\xi ^2})
\end{equation}
is non-degenerate.
We insert Eq.(\ref{eqn:6.5}) into Eq.(\ref{eqn:3.14}), evaluate the integral and then sum the resulting binomial series. We arrive at:
\begin{equation}
\label{eqn:6.6}
\begin{gathered}
  {b_{2n + 1}}\, = \,0,\,\,\,\,\,\,\,\,\,n = 0,1,2,... \hfill \\
  {b_{2n}}\, = \,\frac{1}{{\sqrt {2\pi } }}\frac{{{{( - 1)}^n}}}{{{2^{2n}}}}\,\,\frac{{\sqrt {(2n)!} }}{{n!}},\,\,\,\,\,\,\,\,\,n = 0,1,2,...\, \hfill \\ 
\end{gathered} 
\end{equation}
From this equation we obtain:
\begin{equation}
\label{eqn:6.7}
\begin{gathered}
  {T_{2n + 1}}\, = \,0,\,\,\,\,\,\,\,\,\,n = 0,1,2,... \hfill \\
  {T_{2n}}\, = \,\frac{1}{{\sqrt {8\pi } }}\frac{{{{( - 1)}^n}}}{{{2^{2n}}}}\,\,\frac{{\sqrt {(2n)!} }}{{n(n!)}},\,\,\,\,\,\,\,\,\,n = 1,2,...\, \hfill \\ 
\end{gathered} 
\end{equation}
The condensate wave-function is now given by:
\begin{equation}
\label{eqn:6.8}
\tilde \psi (\xi )\, = \,\sqrt N \{ \,\,{\chi _{\alpha (0)}}(\xi ) - \,\frac{1}{{\sqrt {8\pi } }}\tilde g_1\,N\sum\limits_{n = 1}^\infty  {\frac{{{{( - 1)}^n}}}{{{2^{2n}}}}\,\,\frac{{\sqrt {(2n)!} }}{{n(n!)}}{\chi _{2n}}(\xi )} \,\}  + \,O({(N\tilde g_1)^2})
\end{equation}
This reproduces the results in references \cite{Trallero,kivshar,Yu}.

We determine ${S^{(1)}}$ from Eq.(\ref{eqn:4.3}):
\[{S^{(1)}} = \,\sum\limits_{n = 1}^\infty  {|{T_{2n}}{|^2} = \,\frac{1}{{8\,\pi }}\,\,\sum\limits_{n = 1}^\infty  {\frac{1}{{{2^{4n}}}}\,\,\frac{{(2n)!}}{{{n^2}{{(n!)}^2}}}} }. \]

Using the inequality:
\[(2n)! \leq \,{2^{2n}}\,{(n!)^2},\]
we obtain an upper bound on $S^{(1)}$,
\[{S^{(1)}} \leq \,\frac{1}{{8\,\pi }}\,\sum\limits_{n = 1}^\infty  {\,\,\frac{{{{(\frac{1}{4})}^n}}}{{n^2\,}}\, = \,} \frac{1}{{8\,\pi }}\,g_2(\frac{1}{4}),\]
where
${g_2}\,(\frac{1}{4})$ is the Boson function of order 2 evaluated at, $z = \frac{1}{4}$.
since 
\[{g_2}\,(\frac{1}{4})\, < \,\frac{1}{3},\]
we obtain
\[{S^{(1)}} \, \sim\, \frac{1}{{24\,\pi }}.\]

For the experimental results in \cite{1DBEC} with ${}^{23}Na$, we have
\[{m_b} \simeq 3.8194 \times {10^{ - 26}}Kg,\,\,\,\,a = \,2.8 \times {10^{ - 9}}m\,,\,\,\,\,N = \,3 \times {10^6}\,particles,\]
$${\omega _z} = \sqrt 2 \,{\omega _y} = \,2\,{\omega _x} \equiv 2\omega  = 2\pi  \times 27\,Hz.$$
 Then ${l_c} \simeq 5.7 \times {10^{ - 6}}m$ and $\tilde g_1\ \simeq \,1.65\, \times \,{10^{ - 3}}.$ and ${({{\tilde g}_1}N)^2}{S^{(1)}} = 3.6\,{({10^{ - 4}}\,N)^2}$. Thus for $N \geq {10^4}$ our interpretation is no longer valid because the perturbation theory breaks down for large numbers of particles.
The chemical potential now reads
\begin{equation}
\label{eqn:6.9}
\tilde \mu \, = \,\frac{1}{2} + \,\frac{1}{{\sqrt {2\pi } }}\tilde g_1\,N\, + O({(N\tilde g_1)^2})
\end{equation}
and in the dimension-full units
\begin{equation}
\label{eqn:6.10}
\mu \, = \,\frac{D}{2}\hbar \omega  + \,{(\frac{{{m_b}\omega }}{{2\pi \hbar }})^{D/2}}\,g_1\,N\, + O({(N\,g_1)^2}))
\end{equation} 
with $D = 1$.
To obtain the second order correction, we use Eq.(\ref{eqn:5.2}). This gives:
\[{{\tilde \mu }_2} = \, - \,\frac{3}{{4\,\pi }}\,{N^2}\,\sum\limits_{n = 1}^\infty  {\frac{1}{{{2^{4n}}}}\,\,\frac{{(2n)!}}{{n{{(n!)}^2}}}} \]
This reproduces the second order correction given in \cite{Trallero}.
We obtain the particle density by substituting Eq.(\ref{eqn:6.6}) into Eq.(\ref{eqn:5.4}), this leads to:
\begin{equation}
\label{eqn:6.11}
n(\xi )\, = \,N\,[\chi _0^2(\xi ) - \,\frac{1}{{\sqrt {2\pi } }}\tilde g_1\,N\sum\limits_{n = 1}^\infty  {\frac{{{{( - 1)}^n}}}{{{2^{2n}}}}\,\,\frac{{\sqrt {(2n)!} }}{{n(n!)}}{\chi _{2n}}(\xi )} \,{\chi _0}(\xi )] + \,O({(N\tilde g_1)^2})
\end{equation}

Figures 1(a) and 1(b) show respectively, $\tilde \psi (\xi )\,/\sqrt N $ and $n(\xi )/N$ for $N=1000$ and $N=1300$ particles and for $\tilde{g_1}$ calculated using the experimental data in \cite{1DBEC}. The repulsive inter-particle interactions push the Bose particles from the center of the trapping potential, where they are concentrated to the edges of the trap. As a result the condensate wave-function is extended further and its maximum is lowered and broadened.
If we further increase the number of particles we will observe a dip in both curves indicating the breakdown of perturbation theory.

\begin{figure}[h]
\centering
\subfloat[]{\includegraphics[width=5.5cm]{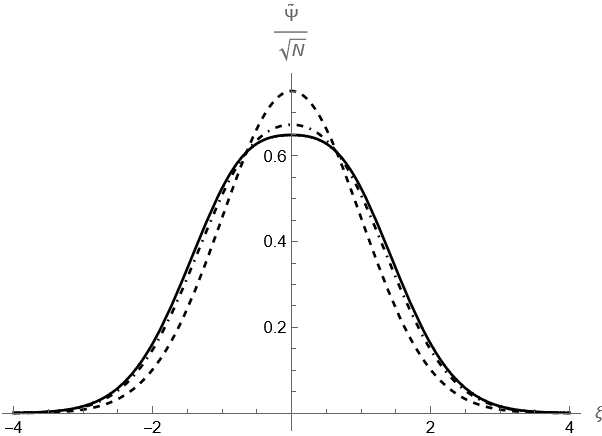}}
\qquad
\subfloat[]{\includegraphics[width=5.5cm]{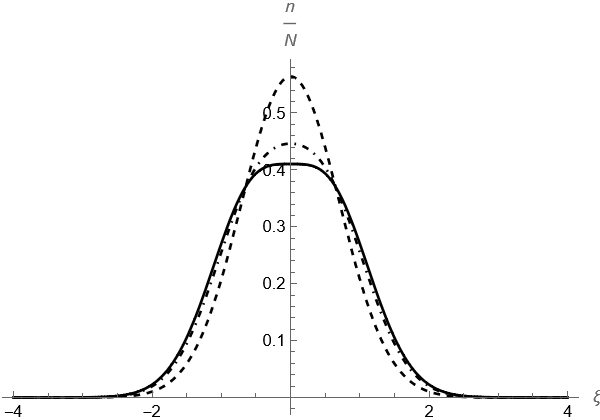}}
\caption*{Fig.1.(a) Condensate wave-function vs $\xi$, (b) particle number density vs $\xi$ for an ideal Bose gas (dashed line), a dilute Bose gas with $N=1000$ particles (dot-dashed line) and $N=1300$ particles (full line).}
\label{Figures6}
\end{figure}

The internal energy of the condensate is given by Eq.(\ref{eqn:5.6}) which now reads:
\begin{equation}
\label{eqn:6.12}
\tilde U(N,\,\omega ) = \,N\,[\frac{1}{2} + \,\frac{1}{{\sqrt {8\pi } }}\tilde g_1\,N\,] + O({(N\tilde g_1)^2})
\end{equation}
or in the dimension-full units
\begin{equation}
\label{eqn:6.13}
U(N,\,\omega ) = \,N\,[\frac{D}{2}\hbar \omega  + \,\frac{1}{2}\,{(\frac{{{m_b}\omega }}{{2\pi \hbar }})^{D/2}}\,g_D\,N\,\,] + O({(N\,g_D)^2})),
\end{equation}
with $D = 1.$
\\We next calculate the mean-square radius of the condensate using Eq.(\ref{eqn:5.11})
\begin{equation}
\label{eqn:6.14}
{\xi _0} = \,\sqrt {N{M_{0,0}}} \,\,[\,1\, - \,\frac{1}{{{M_{0,0}}}}\tilde g_1\,N\,\,\sum\limits_{n = 1}^\infty  {\,{T_n}} \,{M_{n,0}}+ O({(N\tilde g_1)^2})\,],
\end{equation}
where
\[{M_{n,0}} = \,\int {d\xi \,{\xi ^2}\,} \frac{1}{{\sqrt {{2^n}\,n!\,\sqrt \pi  } }}\,\,\exp ( - \frac{1}{2}{\xi ^2})\,{H_n}(\xi )\,{\chi _0}(\xi ).\,\]
Using the recursion formulas for the Hermite polynomials \cite{Lebedevspecialfunctions,Bellspecialfunctions} we obtain
\begin{equation}
\label{eqn:6.15}
{M_{n,0}} = \,(n + \frac{1}{2})\,{\delta _{n,0}} + \,\frac{1}{2}\,\sqrt {n(n - 1)} \,{\delta _{n,2}}.
\end{equation}
Inserting this back into Eq.(\ref{eqn:6.14}) we get:
\begin{equation}
\label{eqn:6.16}
{\xi _0} = \,\sqrt {\frac{N}{2}} \,(\,1\, + \,\frac{1}{{4\sqrt {2\pi } }}\tilde g_1\,N\,) + O({(N\tilde g_1)^2}).
\end{equation}
The inter-particle interactions increase the radius of the condensate.

\section{The two-dimensional isotropic harmonic potential}
We next consider the GP equation in two dimensions with the confining potential
\begin{equation}
\label{eqn:7.1}
{V_{ext}}(x,y) = \frac{1}{2}{m_b}\,{\omega ^2}\,({x^2} + {y^2}).
\end{equation}
In this case the two-dimensional coupling constant is given by \cite{dion}
\[{g_2} = g\,\sqrt {\frac{{{m_b}\,{\omega _z}}}{{2\,\pi \,\hbar }}} ,\]
and 
\begin{equation}
\label{eqn:g2D}
{{\tilde g}_2} = \frac{{4\,\pi \,a}}{{{l_c}}}\sqrt {\frac{{{\gamma _z}}}{{2\pi }}} , 
\end{equation}

where,
\[{l_c} = \,\,\sqrt {\frac{\hbar }{{{m_b}\,\omega }}} ;\,\,\,\,\,\,\,{\gamma _z} = \frac{{{\omega _z}}}{{{\omega}}}.\]

Because of the rotational symmetry of the problem we use plane polar coordinates $\vec \xi  = \,(\xi ,\varphi ),$ where 
\[\xi  = \,\sqrt {\xi _x^2\, + \,\xi _y^2\,} \,\,\,\,\,\,\,\,\,\,\,\,0 \leq \,\,\xi \, < \,\infty ;\,\,\,\,\,\varphi  = {\tan ^{ - 1}}(\frac{{{\xi _y}}}{{{\xi _x}}})\,\,\,\,\,\,\,\,\,\,\,\,0\,\, \leq \varphi  < 2\,\pi \]
Substituting Eq.(\ref{eqn:7.1}) into Eq.(\ref{eqn:3.1}), we obtain the eigenvalue problem for the two-dimensional isotropic harmonic oscillator with eigenvalues:
\begin{equation}
\label{eqn:7.2}
{\lambda _{n,\,m}}\, = \,2n + |m| + 1;\,\,\,\,\,\,\,\,\,\,n = 0,\,1,\,2,\,...;\,\,\,\,\,\,\,\,\,\,m = 0,\,\, \pm 1,\, \pm 2,\,...,
\end{equation}
and eigenfunctions
\begin{equation}
\label{eqn:7.3}
{\chi _{n,m}}(\xi ,\,\varphi )\, = \,{A_{n,\,m}}\,{\xi ^{|m|}}\,\,\exp ( - \frac{1}{2}{\xi ^2})\,L_n^m({\xi ^2})\,\exp (i\,m\,\varphi ),\,\,\,\,\,\,\,\,\,
\end{equation}
where 
\begin{equation}
\label{eqn:7.4}
{A_{n,\,m}}\, = \,\sqrt {\frac{{n!}}{{\pi \,\Gamma (n + m + 1)}}}, \,\,\,\,
\end{equation}
and $L_n^m(x)$ is the associated Laguerre polynomials of order $(n,m)$.
The associated Laguerre polynomials $L_n^k(x)$, where $k$ can be a non-integer, is defined by \cite{Lebedevspecialfunctions,Bellspecialfunctions}:
\begin{equation}
\label{eqn:7.5}
L_n^k(x)\, = \,\frac{{{e^x}\,{x^{ - k}}\,}}{{n!\,}}\,\frac{{{d^n}}}{{d{x^n}}}({e^{ - x}}\,{x^{n + k}}),
\end{equation}
and has the series representation:
\begin{equation}
\label{eqn:7.6}
L_n^k(x)\, = \,\sum\limits_{r\, = \,0}^n {{{( - 1)}^r}} \,\frac{{\Gamma (n + k + 1)}}{{r!\,(n - r)!\,\Gamma (k + r + 1)}}\,{x^r},
\end{equation}
and the generating function:
\begin{equation}
\label{eqn:7.7}
\frac{1}{{{{(1 - t)}^{k + 1}}}}\,\,\exp ( - \,\,\frac{{x\,t}}{{1 - t}}) = \,\sum\limits_{n\, = \,0}^\infty  {L_n^k(x)} \,{x^n}.
\end{equation}
The ground-state eigenfunction,
\begin{equation}
\label{eqn:7.8}
{\chi _{0,0}}(\xi ,\varphi )\, = \frac{1}{{\sqrt \pi  }}\,\,\exp ( - \frac{1}{2}{\xi ^2}),
\end{equation}
is non-degenerate. We determine the coefficients ${b_{n,m}}\,$ by inserting Eq.(\ref{eqn:7.3}) and Eq.(\ref{eqn:7.8}) into Eq.(\ref{eqn:3.14}) and evaluating the integral using the series representation Eq.(\ref{eqn:7.6}). This gives
\begin{equation}
\label{eqn:7.9}
{b_{n,m}}\, = \,\frac{1}{{\pi \,{2^{n + 1}}}}\,{\delta _{m,0}};\,\,\,\,\,\,\,\,\,\,n = 0,\,1,\,2,\,...;\,\,\,\,\,\,\,\,\,\,m = 0,\,\, \pm 1,\, \pm 2,\,...\,.
\end{equation}
From this expression we obtain:
\begin{equation}
\label{eqn:7.10}
{T_{n,m}}\, = \,\frac{1}{{4\,\pi \,}}\frac{1}{{{2^n}\,n}}\,{\delta _{m,0}};\,\,\,\,\,\,\,\,\,\,n =\,1,\,2,\,...;\,\,\,\,\,\,\,\,\,\,m = 0\,\pm 1,\, \pm 2,\,...,\,.
\end{equation}
The condensate wave-function is then
\begin{equation}
\label{eqn:7.11}
\tilde \psi (\xi )\, = \,\sqrt N \{ \,\,{\chi _{0,0}}(\xi ) - \,\frac{1}{{4\,\pi }}\,\tilde g_2\,N\sum\limits_{n = 1}^\infty  {\frac{1}{{{2^n}\,n}}\,\,{\chi _{n,0}}(\xi )} \,\}  + \,O({(N\tilde g_2)^2})
\end{equation}
We determine ${S^{(1)}}$ from Eq.(\ref{eqn:4.3}):
\[{S^{(1)}} = \,\sum\limits_{n = 1}^\infty  {\,\sum\limits_m {|{T_{n,m}}{|^2}}  = \,\frac{1}{{16\,{\pi ^2}}}\,\,\sum\limits_{n = 1}^\infty  {\,\frac{{{{(\frac{1}{4})}^n}\,}}{{{n^2}}}} }  = \,\frac{1}{{16\,{\pi ^2}}}\,{g_2}\,(\frac{1}{4}).\]

This gives 
\[{S^{(1)}}\, \simeq \,\frac{1}{{48\,{\pi ^2}}}\,\]

For the experimental results in \cite{2DBECRb} with ${}^{87}Rb$, we have
\[{m_b} \simeq 1.45 \times {10^{ - 25}}Kg,\,\,\,\omega  = 2\pi \, \times \,20.6 Hz,\,\,\,\,\omega_z  = 2\pi \, \times \,2 KHz,\,\,\,\,a = \,5.3 \times {10^{ - 9}}m\,,\,\,\,\,N = \,6.1 \times {10^4}\,particles.\] 
This gives:
\[\tilde g_2\,N \simeq \,0.11\,N\,\,\,\,\,\,and\,\,\,\,\,\,{({{\tilde g}_2}N)^2}\,{S^{(1)}}\, \simeq \,0.25 \times {({10^{ - 2}}\,N)^2}\]
 Thus for $N \geq {10^2}$ our interpretation is no longer valid because the perturbation theory breaks down for large numbers of particles.
The chemical potential is now given by:
\begin{equation}
\label{eqn:7.12}
\tilde \mu \, = \,1 + \,\frac{1}{{2\pi }}\tilde g_2\,N\, + O({(N\tilde g_2)^2})
\end{equation}
and in the dimension-full units is given by Eq.(\ref{eqn:6.10}) with $D = 2$. Again if we are interested in the second-order correction to the chemical potential we have
\begin{equation}
\label{eqn:7.13}
{{\tilde \mu }_2} = \, - \,\frac{3}{{8\,{\pi ^2}}}\,\ln (\frac{4}{3})\,\,{N^2}\,
\end{equation}
The particle density is now given by:
\begin{equation}
\label{eqn:7.14}
n(\xi )\, = \,N\,[\chi _{0,0}^2(\xi ) - \,\frac{1}{{2\pi }}\tilde g_2\,N\sum\limits_{n = 1}^\infty  {\frac{1}{{{2^{n}}n}}\,\,{\chi _{n,0}}(\xi )} \,{\chi _{0,0}}(\xi )] + \,O({(N\tilde g_2)^2})
\end{equation}

Figures 2.(a) and 2.(b) show respectively, $\tilde \psi (\xi )\,/\sqrt N $ and $n(\xi )/N$ for $N=30$ and $N=40$ particles and for $\tilde{g_2}$ calculated using the experimental data in \cite{2DBECRb}. Again the repulsive inter-particle interaction broadens the peak and shifts it down.

\begin{figure}[h]
\centering
\subfloat[]{\includegraphics[width=5.5cm]{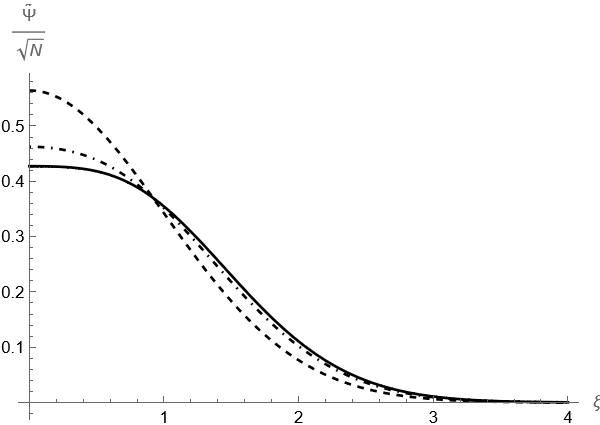}}
\qquad
\subfloat[]{\includegraphics[width=5.5cm]{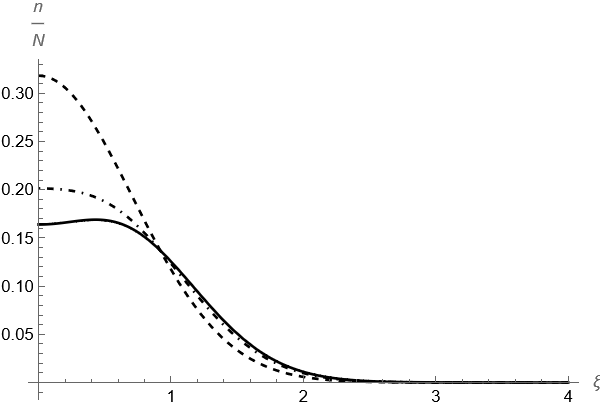}}
\caption*{Fig.2.(a) Condensate wave-function vs $\xi$, (b) particle number density vs $\xi$ for an ideal Bose gas (dashed line), a dilute Bose gas with $N=30$ particles (dot-dashed line) and $N=40$ particles (full line).}
\label{Figures7}
\end{figure}

The internal energy of the gas is obtained from Eq.(\ref{eqn:5.6})
\begin{equation}
\label{eqn:7.15}
\tilde U(N,\,\omega ) = \,N\,[1 + \,\frac{1}{{4\,\pi }}\tilde g_2\,N\,] + O({(N\tilde g_2)^2})
\end{equation}
and in the dimension-full units is given by Eq.(\ref{eqn:6.13}) with $D = 2$.
To calculate the mean square radius of the condensate, we first calculate ${M_{m,n}}$ by substituting Eq.(\ref{eqn:7.3}) into Eq.(\ref{eqn:5.10}). We obtain
\[{M_{(n,m),(0,0)}} = \,{\delta _{m,0}}\,I(n),\]
where \[I(n)\, = \,2\,\int\limits_0^\infty  {d\xi \,{\xi ^3}} \,\exp ( - {\xi ^2})\,{L_n}({\xi ^2}).\]
This integral can be easily evaluated using the generating function for the associated Laguerre polynomials (\ref{eqn:7.7}) with $k = 0.$ This gives:
\[I(0) = \,1,\,\,\,\,\,\,\,\,\,I(1) = \, - 1,\,\,\,\,\,\,\,\,I(n) = 0;\,\,\,\,\,\,n \geq 2\]
Hence the mean square radius of the condensate is:
\[{\xi _0} = \,\sqrt N \,(\,1\, + \,\frac{1}{{8\,\pi }}\tilde g_2\,N\,) + O({(N\tilde g_2)^2}).\]

\section{The three-dimensional isotropic harmonic potential}
We finally consider the GP equation in $D = 3$ with the isotropic harmonic potential
\begin{equation}
\label{eqn:8.1}
{V_{ext}}(x,y,\,z) = \frac{1}{2}{m_b}\,{\omega ^2}\,({x^2} + {y^2} + {z^2}).
\end{equation}
Since the potential ${V_{ext}}(x,y,\,z)$ has a spherical symmetry we use spherical coordinates $\vec \xi  = \,(\xi ,\,\theta ,\,\varphi ),$ where 
\[\begin{gathered}
  \xi  = \,\sqrt {\xi _x^2\, + \,\xi _y^2 + \,\xi _z^2\,} \,\,\,\,\,\,\,\,\,\,\,\,0 \leq \,\,\xi \, < \,\infty ; \hfill \\
  \theta  = {\cos ^{ - 1}}(\frac{{{\xi _x}}}{\xi })\,\,\,\,\,\,\,\,\,\,\,\,0\,\, \leq \theta  < \pi ;\,\,\,\,\,\,\,\varphi  = {\tan ^{ - 1}}(\frac{{{\xi _y}}}{{{\xi _x}}})\,\,\,\,\,\,\,\,\,\,\,\,0\,\, \leq \varphi  < 2\,\pi  \hfill \\ 
\end{gathered} \]
Substituting Eq.(\ref{eqn:8.1}) into Eq.(\ref{eqn:3.1}) we obtain the eigenvalue problem for the isotropic three-dimensional oscillator with eigenfunctions
\begin{equation}
\label{eqn:8.2}
\begin{gathered}
  {\chi _{n,\,l,m}}(\xi ,\,\theta ,\,\,\varphi )\, = \,{A_{n,\,l}}\,{\xi ^l}\,\,\exp ( - \frac{1}{2}{\xi ^2})\,L_n^{l + \frac{1}{2}}({\xi ^2})\,Y_l^m\,(\theta ,\,\,\varphi );\,\,\, \hfill \\
  n = 0,\,1,\,2,\,...;\,\,\,\,\,\,\,\,l = 0,\,1,\,2,\,...;\,\,\,\,\,\,\,\,m = \, - l,\, - l + 1,\,...,\,\,l,\, \hfill \\ 
\end{gathered} 
\end{equation}
where, 
\begin{equation}
\label{eqn:8.3}
{A_{n,\,l}}\, = \,\sqrt {\frac{{2(n!)}}{{\Gamma (n + l + \frac{3}{2})}}} ,\,\,\,\,
\end{equation}
and $Y_l^m\,(\theta ,\,\,\varphi )$ is the spherical harmonic function of order $(l, m)$. The eigenvalues depend only on $n$ and $l$ and are given by:
\begin{equation}
\label{eqn:8.4}
{\lambda _{n,\,l}}\, = \,2n + l + \frac{3}{2};\,\,\,\,\,\,\,\,\,\,n = 0,\,1,\,2,\,...;\,\,\,\,\,\,\,\,\,l = 0,\,1,\,2,\,...\,.
\end{equation}
The ground state eigenfunction, 
\begin{equation}
\label{eqn:8.5}
{\chi _{0,\,0,\,0}}(\xi ,\,\theta ,\,\varphi )\, = {(\frac{1}{\pi })^{\frac{3}{4}}}\,\,\exp ( - \frac{1}{2}{\xi ^2}),
\end{equation}
is non-degenerate. To determine ${b_{n,\,l,\,m}}$ we insert Eq.(\ref{eqn:8.2}) and (\ref{eqn:8.5}) into Eq.(\ref{eqn:3.14}), this gives
\begin{equation}
\label{eqn:8.6}
{b_{n,\,l,\,m}}\, = \,2\,{(\frac{1}{\pi })^{\frac{7}{4}}}\,{A_{n,\,l}}\,{\delta _{l,0}}\,{\delta _{m,0}}\,I(n),
\end{equation}
where,
\[I(n)\, = \,\int\limits_0^\infty  {d\xi \,{\xi ^2}} \,\exp ( - 2\,{\xi ^2})\,L_n^{\frac{1}{2}}({\xi ^2}).\]
We may use the series expansion Eq.(\ref{eqn:7.6}) with $k = 1/2$ to evaluate the integral with the result,
\[I(n)\,\, = \,\frac{1}{{{2^{n + \frac{5}{2}}}\,n!}}\Gamma (n + \frac{3}{2}).\]
Substituting this back into Eq.(\ref{eqn:8.6}), we obtain after some algebra
\begin{equation}
\label{eqn:8.7}
{b_{n,\,l,\,m}}\, = \,\frac{1}{{{2^{2n + 2}}\,n!}}\,{[\,\frac{{(\,2n + 2\,)!}}{{{\pi ^3}\,(\,n + 1\,)}}\,]^{\frac{1}{2}}}\,\,{\delta _{l,0}}\,{\delta _{m,0}};\,\,\,\,\,\,\,\,\,\,\,\,\,\,\,n = 0,\,1,\,2,\,...\,.\,
\end{equation}
Then,
\begin{equation}
\label{eqn:8.8}
{T_{n,\,l,\,m}}\, = \,\frac{1}{{{2^{2n + 3}}\,n\,(n!)}}\,{[\,\frac{{(\,2n + 2\,)!}}{{{\pi ^3}\,(\,n + 1\,)}}\,]^{\frac{1}{2}}}\,\,{\delta _{l,0}}\,{\delta _{m,0}};\,\,\,\,\,\,\,\,\,\,\,\,\,\,\,n = 1,\,2,\,...\,.\,
\end{equation} 
The condensate wave-function now reads:
\begin{equation}
\label{eqn:8.9}
\tilde \psi (\xi )\, = \,\sqrt N \{ \,\,{\chi _{0,\,0,\,0}}(\xi ) - \,{(\frac{1}{{4\,\pi }})^{\frac{3}{2}}}\,\tilde g\,N\sum\limits_{n = 1}^\infty  {\,\frac{1}{{{2^{2n}}\,n\,(n!)}}\,{{[\,\frac{{(\,2n + 2\,)!}}{{\,(\,n + 1\,)}}\,]}^{\frac{1}{2}}}\,\,{\chi _{n,\,0,0}}(\xi )} \,\}  + \,O({(N\tilde g)^2})
\end{equation}
We calculate ${S^{(1)}}$ using the inequality:
\begin{equation}
\label{inequality}
(2n + 1)(2n)! \leq \,{2^{2n + 1}}\,n\,{(n!)^2},
\end{equation}
leading to
\[{S^{(1)}} = \,\sum\limits_{n = 1}^\infty  {\,\sum\limits_{l,\,m} {|{T_{n,\,l,\,m}}{|^2}} }  \leq \,\frac{1}{{16\,{\pi ^3}}}\,\sum\limits_{n = 1}^\infty  {\,\frac{{{{(\frac{1}{4})}^n}\,}}{n}}  = \,\frac{1}{{16\,{\pi ^3}}}\,\ln (\frac{4}{3}).\]

For the experimental results in \cite{3DBECRb} using $^{87}Rb$ atoms, we have
\[{m_b} \sim \,1.45 \times {10^{ - 25}}Kg,\,\,\,{\omega} \sim \,2\pi \times 77.5\,\, Hz,\,\,\,\,a \approx \,5.8 \times {10^{ - 9}}m\,,\,\,\,\,N \sim \,1.2 \times {10^5}\,particles\]
This gives:
\[\tilde g\,N \simeq \,0.61 \times {10^{-1}}\,N\,\,\,\,\,\,\]
The fraction of the Bose particles tunneling to the excited states is now
\[{(\tilde g\,N)^2}\,{S^{(1)}} \sim \,0.32\, \times (10^{-2} N)^2.\]
For $N \geq {10^2}$ our interpretation is no longer valid, this is because the actual expansion coefficient $\tilde g\,N$ is very large and the perturbation theory breaks down.
The chemical potential is given by 
\begin{equation}
\label{eqn:8.10}
\tilde \mu \, = \,\frac{3}{2} + \,{(\frac{1}{{2\pi }})^{\frac{3}{2}}}\tilde g\,N\, + O({(N\tilde g)^2}),
\end{equation}
and in the dimension-full units by Eq.(\ref{eqn:6.10}) with $D = 3$.
The second order correction is given by Eq.(\ref{eqn:5.2}), which now reads:
\begin{equation}
\label{eqn:8.11}
{{\tilde \mu }_2} \simeq \, - \,\frac{{3\,{N^2}}}{{16\,{\pi ^3}}}\,\sum\limits_{n = 1}^\infty  {\,\frac{{(2n + 1)\,(\,2n\,)!}}{{{2^{4n}}\,n\,{{(n!)}^2}}}\,.\,} \,
\end{equation}
Instead of summing this series we will obtain an upper bound on its sum using the inequality Eq.(\ref{inequality}). This gives
\begin{equation}
\label{eqn:8.12}
{{\tilde \mu }_2} \simeq \, - \,\frac{{3\,{N^2}}}{{8\,{\pi ^3}}}\,\sum\limits_{n = 1}^\infty  {\,{{(\frac{1}{4})}^n}\, = \,} \, - \,\frac{{{N^2}}}{{8\,{\pi ^3}}}.
\end{equation}
The particle density now reads
\begin{equation}
\label{eqn:8.13}
n(\xi )\, = \,N\{ \,\,\chi _{0,\,0,\,0}^2(\xi ) - \,{(\frac{1}{{2\,\pi }})^{\frac{3}{2}}}\,\tilde g\,N\sum\limits_{n = 1}^\infty  {\,\frac{{\sqrt {(\,2n + 1\,)!} }}{{{2^{2n}}\,n\,\,(n!)}}\,\,{\chi _{n,\,0,0}}(\xi )} \,{\chi _{0,\,0,\,0}}(\xi )\}  + \,O({(N\tilde g)^2}).
\end{equation}
Figures 3.(a) and 3.(b) show respectively, $\tilde \psi (\xi )\,/\sqrt N $ and $n(\xi )/N$ for $N=30$ and $N=40$ particles and for $\tilde{g}$ calculated using the experimental data in \cite{3DBECRb}. Again the repulsive inter-particle interaction broadens the peak and shifts it down. 

\begin{figure}[h]
\centering
\subfloat[]{\includegraphics[width=5.5cm]{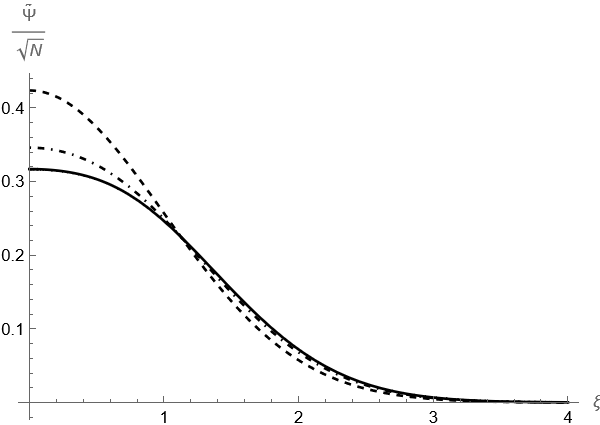}}
\qquad
\subfloat[]{\includegraphics[width=5.5cm]{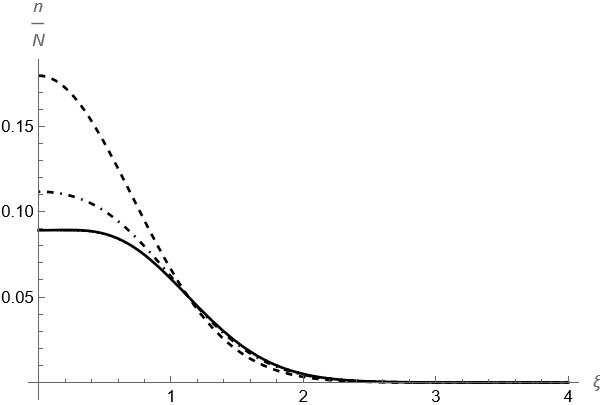}}
\caption*{Fig.3.(a) Condensate wave-function vs $\xi$, (b) particle number density vs $\xi$ for an ideal Bose gas (dashed line), a dilute Bose gas with $N=30$ particles (dot-dashed line) and $N=40$ particles (full line).}
\label{Figures8}
\end{figure}

We calculate the internal energy of the gas using Eq.(\ref{eqn:5.6})
\begin{equation}
\label{eqn:8.14}
\tilde U(N,\,\omega ) = \,N\,[\frac{3}{2} + \,\frac{1}{2}\,{(\frac{1}{{2\,\pi }})^{\frac{3}{2}}}\tilde g\,N\,] + O({(N\tilde g)^2}),
\end{equation}
And in the dimension-full units is given by Eq.(\ref{eqn:6.13}) with $D = 3$.
Finally we calculate the mean-square radius of the condensate by inserting Eqs.(\ref{eqn:8.2}) and (\ref{eqn:8.5}) into Eq.(\ref{eqn:5.10})
\begin{equation}
\label{eqn:8.15}
{M_{(n,\,l,\,m),\,(0,\,0,\,0)}} = \,2\,{(\frac{1}{\pi })^{\frac{1}{4}}}\,{[\,\frac{{2(\,n\,!)}}{{\,\Gamma (\,n + \frac{3}{2}\,)}}\,]^{\frac{1}{2}}}I(n)\,{\delta _{l,0}}\,{\delta _{m,0}}\,,
\end{equation}
where,
\[I(n)\, = \,\frac{1}{2}\,\,\int\limits_0^\infty  {dx\,{x^{\frac{3}{2}}}} \,{\operatorname{e} ^{ - x}}\,L_n^{\frac{1}{2}}(x).\]
Using the generating function Eq.(\ref{eqn:7.7}) with $k= \frac{1}{2}$ we obtain:
\[I(0) = \,\frac{1}{2}\,\Gamma (\frac{5}{2}),\,\,\,\,\,\,\,\,\,I(1) = \, - \frac{1}{2}\,\Gamma (\frac{5}{2}),\,\,\,\,\,\,\,\,I(n) = 0;\,\,\,\,\,\,n \geq 2.\]
This gives 
\begin{equation}
\label{eqn:8.16}
{\xi _0} = \,\frac{3}{2}\sqrt {\frac{N}{\pi }} \,(\,1\, + \,\frac{{\sqrt 2 }}{{16\,\pi }}\tilde g\,N\,) + O({(N\tilde g)^2}).
\end{equation}

\section{Conclusion}
We  have studied the perturbative solution of the Gross-Pitaevskii(GP) equation in the $D$-dimensional space $R^D$ with a general confining potential, $V_{ext}(\vec{r})$. The solution describes the condensate wave-function of a gas of $N$ Bose particles under the influence of the external potential and the two-body inter-particle interactions ${g_D}\,{\delta ^D}(\vec r - \vec r')$. 
\\We obtained the condensate wave-function corrected to first order in, $g_D\,N$, which is the actual expansion parameter. We showed that if the number $N$ of the Bose particles exceeds a certain number, which depends on $g_D$ and the dimension $D$, of the condensate then the perturbation theory breaks down. We calculated the physical parameters; the particle density, the chemical potential, the internal energy and the mean-square radius of the condensate to first order in, $g_D\,N$.\\
We applied the method to the GP equation in $D=1$ with a harmonic potential. The solution represents a cigar-shaped Bose condensate using the experimental data \cite{1DBEC}. We showed that the perturbative solution breaks down if $N$ exceeds $1300$ particles. We also studied the $D=2$ and $D=3$ with a rotationally symmetric and spherically symmetric harmonic potentials respectively. In both cases the maximum number of particles that can be described by the perturbative solution does not exceed $40$ particles for the data in references \cite{2DBECRb,3DBECRb}.\\
This is a major disadvantage of the perturbative approach since it cannot be used to compare with the experimental results which use a number of particles in the range $10^6 \sim 10^7$. However, it sheds light on the nature of the solution and allows us to compute important physical parameters of the system as we have seen.\\
It is very interesting to apply the perturbative method to obtain solutions of the GP equation with more realistic two-body interactions. Also to extend the method to the investigation of the time-dependent GP equation as well as to the study of the excited states of the Bose condensate.

\bibliography{paperall}
\end{document}